\documentclass{emulateapj}
\usepackage{graphicx}
\usepackage{epsfig}
\usepackage{multirow}
\usepackage{hyperref}
\usepackage{xcolor}
\usepackage{natbib}
\newcommand{\ch}{{\it Chandra}\ }
\newcommand{\chandra}{{\it Chandra}\ }

\newcommand{\xmmn}{{\it XMM-Newton}\ }
\newcommand{\nustar}{{\it NuSTAR}\ }
\newcommand{\bepposax}{{\it BeppoSAX}\ }
\begin{document}

\title{A Focused, Hard X-ray Look at Arp 299 with NuSTAR}
\author{A. Ptak\altaffilmark{1,2}, A. Hornschemeier\altaffilmark{1,2}, A. Zezas\altaffilmark{3,4}, B. Lehmer\altaffilmark{2,1}, M. Yukita\altaffilmark{2,1}, D. Wik\altaffilmark{2,1}, V. Antoniou\altaffilmark{4}, 
M. K. Argo\altaffilmark{5}, L. Ballo\altaffilmark{6}, K. Bechtol\altaffilmark{7},  S. Boggs\altaffilmark{8}, R. Della Ceca\altaffilmark{6}, F. E. Christensen\altaffilmark{9}, W. W. Craig\altaffilmark{8,10}, C. J. Hailey\altaffilmark{11}, F. A. Harrison\altaffilmark{12}, R. Krivonos\altaffilmark{8}, T. J. Maccarone\altaffilmark{13}, D. Stern\altaffilmark{14},
M. Tatum\altaffilmark{15,1}, T. Venters\altaffilmark{1}, W. W. Zhang\altaffilmark{1}}

\altaffiltext{1}{NASA Goddard Space Flight Centre, Code 662, Greenbelt, MD 20771, USA} 
\altaffiltext{2}{The Johns Hopkins University, Homewood Campus, Baltimore, MD 21218, USA}
\altaffiltext{3}{Physics Department, University of Crete, Heraklion, Greece}
\altaffiltext{4}{Harverd-Smithsonian Center for Astrophysics, 60 Garden Street, Cambridge, MA 02138, USA}
\altaffiltext{5}{Jodrell Bank Centre for Astrophysics, The University of Manchester, Oxford Rd, Manchester M13 9PL, UK}
\altaffiltext{6}{Osservatorio Astronomico di Brera (INAF), via Brera 28, I-20121, Milano (Italy)}
\altaffiltext{7}{Kavli Institute for Cosmological Physics, Chicago, IL 60637, USA}
\altaffiltext{8}{U.C. Berkeley Space Sciences Laboratory, Berkeley, CA, USA}
\altaffiltext{9}{National Space Institute, Technical University of Denmark, Copenhagen, DK}
\altaffiltext{10}{Lawrence Livermore National Laboratory, Livermore, CA, USA}
\altaffiltext{11}{Columbia University, New York, NY, USA}
\altaffiltext{12}{Caltech Division of Physics, Mathematics and Astronomy, Pasadena, USA}
\altaffiltext{13}{Department of Physics, Texas Tech University, Lubbock, TX, 79409, USA}
\altaffiltext{14}{Jet Propulsion Laboratory, California Institue of Technology, Pasadena, CA 91109, USA}
\altaffiltext{15}{NASA Postdoctoral Program Fellow}

\begin{abstract}
We report on simultaneous observations of the local starburst system
Arp~299 with \nustar\ and {\it Chandra}, which provides the first resolved images
of this galaxy up to energies of $\sim 45$ keV.
Fitting  the 3--40~keV spectrum reveals a column density of $N_{\rm H} \sim 4 \times 10^{24}$~cm$^{-2}$, characteristic of a Compton-thick AGN, and a $10-30$ keV luminosity of $1.2\times 10^{43}$ ergs s$^{-1}$.  The hard X-rays detected by \nustar above 10~keV are centered on the western nucleus, Arp 299-B, which previous X-ray observations have shown to be the primary source of neutral Fe-K emission.  Other X-ray sources, including Arp~299-A, the eastern nucleus which is also thought to harbor an AGN, as well as X-ray binaries, contribute
$\lesssim 10\%$ to the 10--20~keV emission from the Arp 299 system. 
  The lack of significant emission above 10~keV other than that attributed to Arp 299-B suggests that: a) any AGN in Arp 299-A must be heavily obscured ($N_{\rm H} > 10^{24}$ cm$^{-2}$) or have a much lower luminosity than Arp 299-B and b) the extranuclear X-ray binaries have spectra that cut-off above $\sim 10$ keV.  Such soft spectra are characteristic of ultraluminous X-ray (ULX) sources observed to date by {\it NuSTAR}.
\end{abstract}

\keywords{galaxies: active --- galaxies: individual (Arp 299) --- galaxies: starburst --- X-rays}

\section{Introduction\label{introduction}}
 Interactions and mergers are key mechanisms for driving gas into the nuclear regions of galaxies, which in turn are believed to fuel both active galactic nuclei (AGN) and circumnuclear star formation.  However, it is only in the latest stages of an interaction when the AGN is uncovered from its surrounding material and can be viewed in the optical and/or soft X-ray bands \citep[e.g.,][]{Hopkins:2006p2661}.
 Therefore, observations in hard X-rays ($E >10$ keV) are fundamental for identifying the large number of AGN in galaxy mergers that are still enshrouded in dense circumnuclear material.  Indeed, binary AGN have been detected in several galaxies now, with candidates often selected from spectroscopic surveys 
\citep[e.g.,][]{Comerford:2009fo,Liu:2012jw,Cotini:2013fw}.
However, some of the most convincing evidence of binary AGN, including highly obscured AGN, comes from the detection of Fe-K X-ray lines from both nuclei,
\citep[e.g., NGC 6240 and NGC 3393 in ][respectively]{Komossa:2002js,Fabbiano:2011fo}. Identification of such systems and measurement of the conditions of the circumnuclear obscuring material and the intrinsic luminosity of the AGN are important for understanding the growth of supermassive black holes, since it is generally believed that this growth often takes place in galaxy mergers and is heavily obscured  
\citep[e.g.,][]{Fabian:1998es,2014arXiv1403.7531S}.
 
Arp 299 is an excellent local laboratory for studying ``binary'' AGN in merging galaxies. With $L_{\rm IR} = 8 \times 10^{11} L_{\rm \odot}$ the Arp 299 system (D = 44 Mpc) has one of the highest IR luminosities in the nearby universe, comparable in luminosity to archetypal dual AGN merger galaxy NGC 6240 and almost luminous enough to be classified as an ultraluminous infrared galaxy (ULIRG; galaxies with $L_{\rm IR} > 10^{12} L_{\rm \odot}$).  As is the case for most ULIRGs, Arp 299 is a merger system, with the eastern galaxy referred to as Arp 299-A and the western galaxy being Arp 299-B.\footnote{In the literature the western nucleus is often denoted as NGC 3690 and the eastern nucleus as IC 694. However this is a mis-identification and both nuclei are NGC 3690 with IC 694 being a galaxy $\sim 1\arcmin$ to the northwest, and well outside, of the Arp 299 system.}

\begin{figure}
\epsfig{file=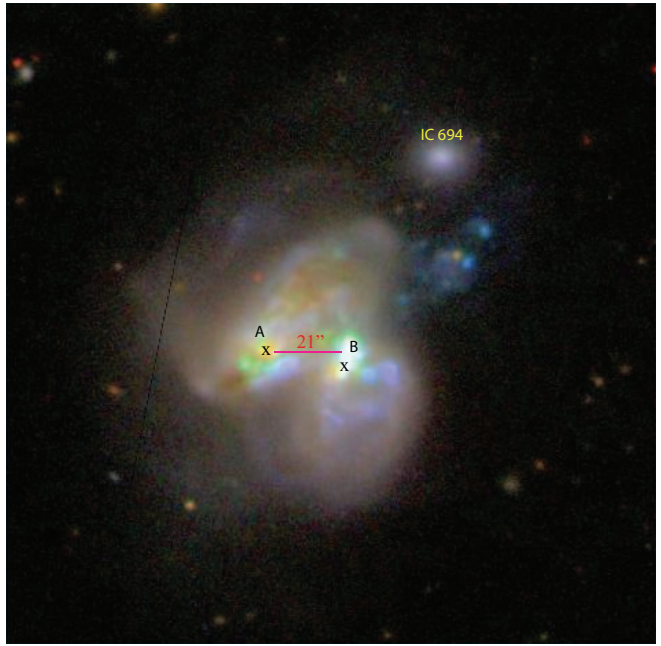,width=\linewidth}
\caption{\label{sdss} SDSS image of Arp 299 downloaded from the SDSS DR7, with the \chandra positions of the two nuclei from \citet{2003ApJ...594L..31Z} marked. IC 694 is also identified.}
\end{figure}

Figure \ref{sdss} shows the SDSS image of Arp 299 with the positions of the two nuclei marked based on their potential X-ray counterparts 
\citep[sources 6 and 16 from][separated by $\sim 21\arcsec = 4.2$ kpc]{2003ApJ...594L..31Z}.   
\citet{DellaCeca:2002hc} detected hard X-ray emission from Arp 299 with \bepposax and showed that the system hosts a heavily obscured AGN ($N_{\rm H} \sim 2 \times 10^{24}\ \rm cm^{-2}$). 
Using short \chandra and \xmmn observations, \citet{2003ApJ...594L..31Z} and
\citet{2004ApJ...600..634B} claimed that Arp 299 exhibits evidence for a binary AGN, with hard X-ray spectra and Fe-K$\alpha$ line emission being detected from both nuclei. However, the Fe-K$\alpha$ lines detected from the two nuclei differ in that Arp 299-B exhibits neutral Fe-K, as is commonly observed in AGN, while Arp 299-A exhibits ionized Fe-K. which is not as common for AGN.  LIRGs (luminous infrared galaxies with
 $L_{\rm IR} > 10^{11}~L_{\odot}$) and ULIRGs are more frequently ionized Fe-K emitters \citep{Iwasawa:2009p4997}.   While the Fe-K emission may imply that both nuclei are harboring AGN, the neutral Fe-K from Arp 299-B is more consistent with that nucleus being the heavily obscured AGN detected by \bepposax and the ionized Fe-K in Arp 299-A could have significant contributions from star formation.  While some ionized Fe-K in Arp 299-A could be due to AGN emission, e.g., a photoionized accretion disk, it is also possible that some ionized Fe-K is tied to star formation, namely high-mass X-ray binaries (HMXBs) and supernova remnants \citep[e.g., see][in the case of NGC 253]{Mitsuishi:2011cp}.

Interestingly, 
\citet{2010A&A...519L...5P} argue that Arp 299-A harbors an AGN based on the detection of a strong, flat-spectrum radio source with a jet-like morphology ($L_{1.7-8.4\rm \ GHz} \sim 2 \times 10^{37}\ \rm ergs\ s^{-1}$).  Recently   
\citet{AlonsoHerrero:2013ct}
reported on {\it Spitzer} mid-IR spectral modeling of the nuclear sources in Arp 299, and similarly found evidence of highly obscured AGN in both nuclei.  They found that the AGN in Arp 299-A must be both more obscured and much less luminous than the AGN in Arp 299-B.  Unfortunately, the limited bandpass of the \chandra and \xmmn observations does not allow us to measure the column densities of the obscuring material and the intrinsic luminosities of heavily obscured AGN. 

\nustar \citep{Harrison:2013iq}, the first X-ray satellite capable of focusing X-rays above 12 keV, can improve our understanding of the Arp 299 system in several dramatic ways.  First, while Arp 299 was detected to 30--35 keV in 2001 by \bepposax 
\citep{DellaCeca:2002hc}, the FOV of the (non-imaging) PDS detector is $\sim 60\arcmin$ and there is no way to disentangle the $E>10$ keV emission from Arp299-A and Arp299-B.  While \nustar has a $\sim 1'$ half-power diameter (HPD) point-spread function (PSF), the core of the PSF (FWHM $\sim 18\arcsec$) is sharp enough to determine which of the two nuclei, if either, is dominating the hard X-ray luminosity of Arp 299.  Second, while the X-ray binary population probed by \chandra in past observations is not fully resolvable by {\it NuSTAR}, if these binaries are emitting prominently above 10 keV then this should result in detectable extended emission.  In NGC 253, the combined imaging power of \chandra and \nustar enabled the determination of the energy balance between the central supermassive black hole (SMBH) and the nuclear X-ray binary population 
\citep{Lehmer:2013bk,Wik:2014wt}.

Arp 299 is part of a \nustar program to observe starburst galaxies jointly with other X-ray satellites.
The overarching goal is to characterize the X-ray binary populations of these galaxies, particularly above 10 keV.  In addition, \nustar will set the tightest X-ray constraints to date on the presence of highly obscured AGN that might be present in these starbursts (Arp 299 is the only case with strong existing evidence for an AGN), and to search for diffuse hard X-ray emission that might be due to inverse-Compton (IC) scattering of IR photons (see Wik et al. 2014b).  The detection of an IC component, or a tight upper limit, might help distinguish between hadronic and leptonic cosmic ray models for the production of gamma ray emission in starburst galaxies 
\citep[see, e.g.,][]{Lacki:2013hz}.  We assume $H_0 = 70 \rm km\ s^{-1}\ Mpc^{-1}$.

\section{Observations}
\nustar and \chandra observed Arp 299 simultaneously on 2013 March 13. The \nustar observation was split in two exposures (on the same day, separated by 4 hours) and the data were reduced using HEADAS 6.14 and NuSTARDAS 1.2.  Standard \nustar processing routines (called from the pipeline command nupipeline) were run on the datasets to produce calibrated level 2 events lists and auxiliary data.  The resulting good exposure times were 9.4 and 59.8 ks for the two \nustar datasets.  There are two focal plane modules on \nustar, FPMA and FPMB, each consisting of four chips.  We found no significant spatial offset between the sky coordinates computed for the two detectors, and so for image analysis we summed the FPMA and FPMB images.  We extracted spectra from a 1\arcmin\ radius region centered on the peak of the X-ray emission from Arp 299 and generated spectral responses (using nuproducts) for each observation and detector.  Both the \nustar and \chandra spectra were binned to 30 counts per bin to allow for $\chi^2$ fitting.  We used $\chi^2$ minimization for initial fitting of the spectra with simple models (i.e., a power-law) to assess the quality of the data and the need for more complex models. However, 
for complex models we used the C-statistic, with the spectra binned to 3 counts per bin (which improves the robustness of using the C-statistic, based on simulations by the \nustar science team), since the C-statistic is more appropriate for Poisson-distributed data.  Errors are given assuming $\Delta C=4.6$, or 90\% confidence for two interesting parameters.

\begin{figure}
\epsfig{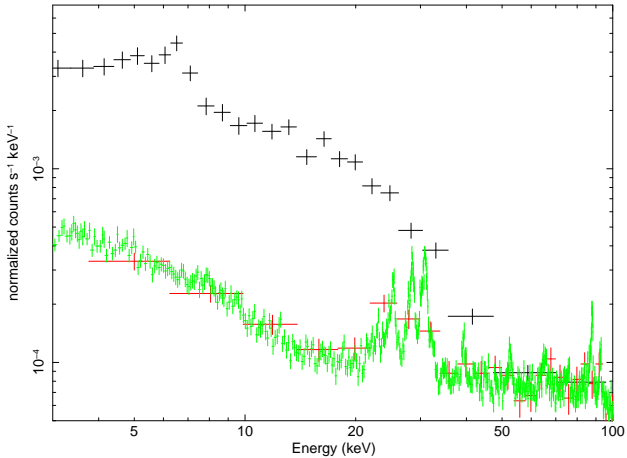}
\caption{\label{bgds} \nustar total (black points) counts within the 1\arcmin\ source aperture for observation 2 (with the highest exposure time) plotted along with the `local' background inferred from source-free regions adjacent to the source (red points) and a model of the background (green points), plotted for FPMA. The local and modeled backgrounds agree well, and comparison with the total counts demonstrates that the Arp 299 system is detected out to $\sim 40-50$ keV.}
\end{figure}

We selected nearby source-free regions to extract the background, 
chosen to lie on the same chip as the source since the intensity of the instrumental background at $\sim 22-25$ keV varies from chip to chip \citep{Wik:2014vv}.  
As an independent check on the background we computed the background spectra expected in the source regions using the procedure described in 
\citet{Wik:2014vv}.
Briefly, this approach fits the separate components of the detector, unfocused X-ray and focused X-ray backgrounds from a model to source-free regions.  Figure \ref{bgds} shows the total source counts in the 1\arcmin\ source region for observation 2 along with the two background estimates, which agree very well.   
 In the spectral analysis discussed below we assumed the local background since this inherently includes statistical error.

The \chandra data were reduced with {\it CIAO} v. 4.6 and CALDB v 4.6. The net exposure time of the \chandra Arp 299 observation was 90.0 ks. Detailed analysis of the \chandra data will be presented in Zezas et al. (in prep.).  \chandra and \nustar overlap in the $3-7$ keV energy band and the \chandra image in this band is shown in Figure \ref{chnuimgs} (right panel).  Here we intend to focus on the joint analysis of the \nustar and \chandra data.  Therefore spectra were extracted and responses were generated for a 1\arcmin\ region corresponding to the \nustar source region to allow for simultaneous fitting with the \nustar spectra.  Responses were generated for the same region.
The same 1\arcmin\ region with the western nucleus excluded was also extracted and fit independently. As shown below, this nucleus dominates the hard X-ray emission; therefore, this analysis allows us
to assess the spectrum of the extranuclear contributions to the {\it NuSTAR} spectrum below 10 keV.

\section{Results}
\subsection{Spatial Results}
\begin{figure*}
\epsfig{file=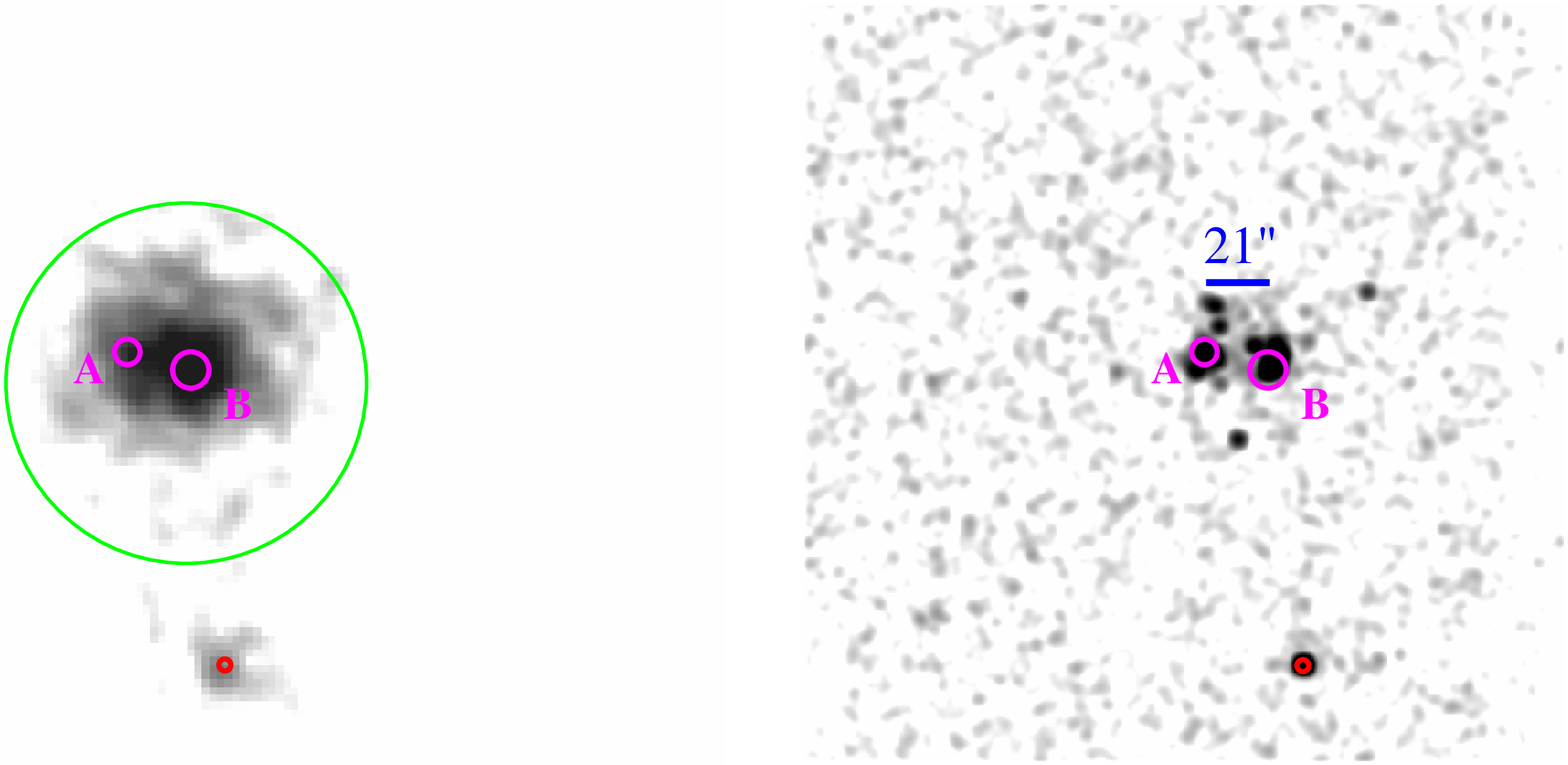,width=\linewidth}
\caption{\label{chnuimgs} Adaptively smoothed images in the 3--7 keV band from the simultaneous observation of Arp 299 from \nustar (left) and \ch (right). The same sky region is shown in both images.  In the \nustar image both detectors and observations are combined. Both images are shown with a logarithmic intensity scale and marked (with red circles) with the positions of Arp 299 A and B, and a serendipitous source detected to the south of Arp 299.  The green circle in the \nustar image is 1\arcmin\ in radius and shows the region used for extracting spectra.}
\end{figure*}
Figure \ref{chnuimgs} (left panel) shows the 3--7 keV \nustar image, with the nuclei of the two Arp 299 merger galaxies A and B marked, as well as a serendipitous source detected 1.6\arcmin\ to the south of Arp 299 that is considerably fainter than the Arp 299 system.  The serendipitous source demonstrates that the \chandra and \nustar frames are fairly well aligned, except for a few arcsecond shift in declination.  For spatial analysis we proceed utilizing just the 2nd, deeper observation to avoid uncertainties due to the alignment of the two observations. 

\begin{figure}
\epsfig{file=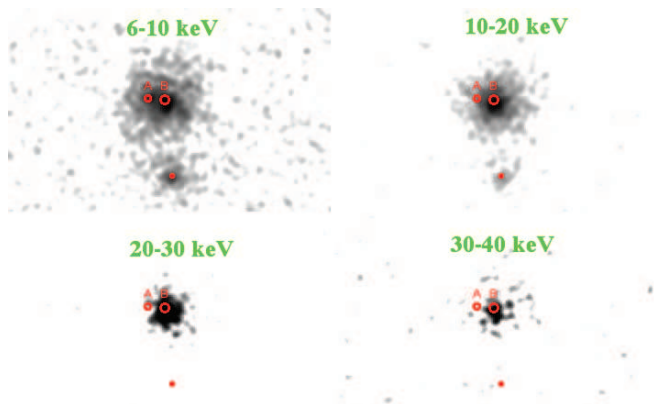,width=\linewidth}
\caption{\label{imgsbyE} The 6--10, 10--20, 20--30 and 30--40 keV \nustar images of Arp 299 from the 2nd observation (FPMA + FPMB, without background subtraction).  The red circles mark the \chandra positions of Arp 299-A, Arp 299-B and the serendipitous source as in Figure \ref{chnuimgs}.}
\end{figure}

\begin{figure}
\epsfig{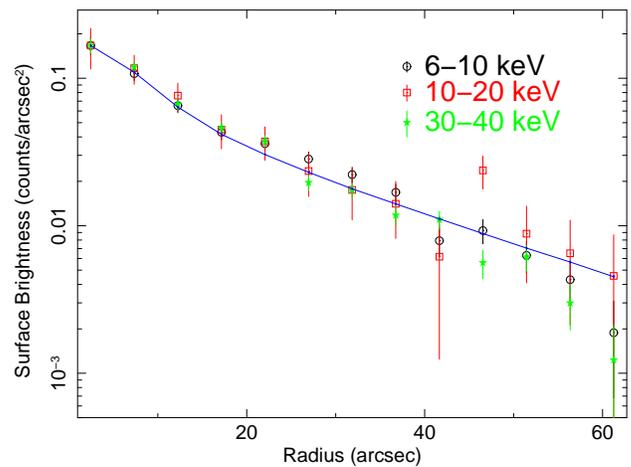}
\caption{\label{radprof} Radial profiles in the 6--10 keV (circles), 10--20 keV (squares) and 30--40 keV energy ranges (stars).  Also shown is the PSF model convolved with a 2'' Gaussian as a solid line.}
\end{figure}

Figure \ref{imgsbyE} shows the 6--10, 10--20, 20--30 and 30--40 keV \nustar images of Arp 299 (note that the PSF is largely independent of energy).  The \nustar image of the Arp 299 system below $\sim 10$ keV is extended, evidently due in large part to both nuclei contributing to the emission as was known to be the case from previous \ch and \xmmn observations.  Above 10 keV the flux becomes increasingly symmetrical, point-like and centered on Arp 299-B.   In Figure \ref{radprof} we show radial profiles centered on Arp 299-B, in the energy ranges 6--10 keV, 10--20 keV and 30--40 keV as well as the profile resulting from a PSF model image.\footnote{PSF images were generated from calibration data using the IDL routine CreateEffPSF.pro;  we used the same approximately on-axis PSF model for the sources belonging to the Arp 299 system and the PSF is largely independent of energy.}  Except for a small excess around 20--30\arcsec\ in the 6--10 keV radial profile, all three profiles are consistent with the PSF.

\begin{figure}
\epsfig{file=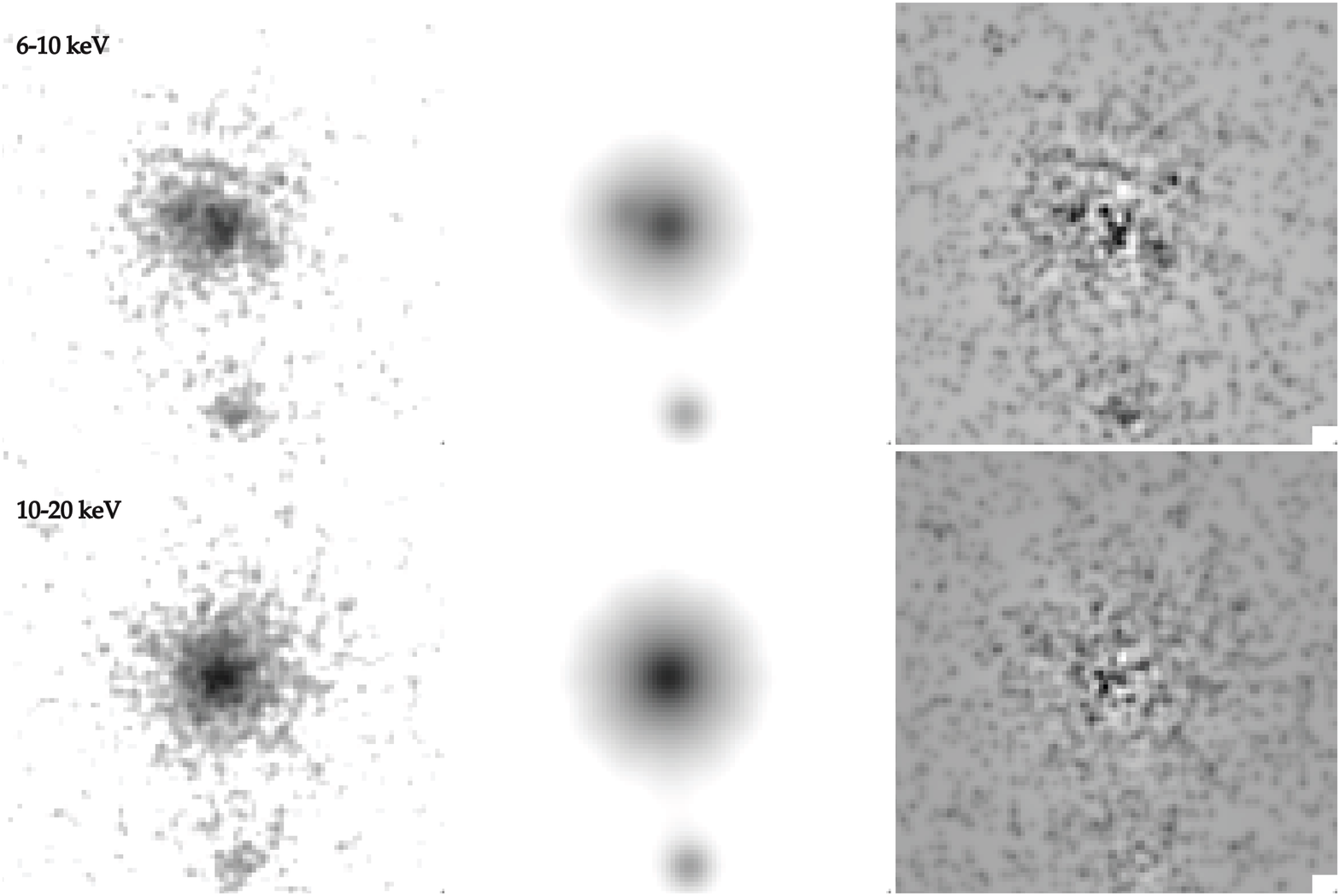,width=1.0\linewidth}
\caption{\label{spatfitimgs} Spatial fitting of the 6--10 keV (top) and 10--20 keV (bottom) \nustar images, from observation 2 (FPMA + FPMB). All images have been smoothed with a 2\arcsec\ Gaussian for display purposes and the data and model images are shown with logarithmic scaling.  The panels are data (left), model (middle) and data  - model residuals (right).  The residuals ranged from -3.3 -- 4.7 counts/pixel in the 6-10 keV fit and -3.9 -- 8.1 counts/pixel in the 10-20 keV fit (residuals in the smoothed images shown are -0.8 -- 1.6 and -1.2 -- 2.6, respectively).}
\end{figure}

To quantify the relative contributions of Arp 299-B and the remaining Arp 299 components we modeled the 6--10 keV image with circular Gaussian images convolved with PSF models. Note that for a circular Gaussian model, $\sigma$ corresponds to $\sim$ the half-power radius (of the intrinsic emission, i.e., prior to forward-folding).  We fit a $100 \times 100\ (4.1\arcmin\ \times 4.1\arcmin$) pixel image with two sources which is sufficient to model the Arp 299 flux. A third source was used to fit the southern serendipitous source  (a separate PSF generated for the serendipitous source was slightly elongated due to its larger off-axis angle). The best-fit parameters (source position, $\sigma$ and normalization) were determined using the Marquardt-Levenberg algorithm with the C-statistic.  We included background images generated by the algorithm described in \citet{Wik:2014vv} in the fits.  Using elliptical Gaussians with free rotation angles did not result in a significant improvement to the fit.  The fit parameters and errors 
are listed in Table \ref{spattab} and the image, best-fit model, and residuals are shown in Figure  \ref{spatfitimgs}.  The offset between the \chandra and \nustar positions for the southern source is 1.9\arcsec\ in R.A. and 3.6\arcsec\ in Declination, consistent with the typical attitude reconstruction accuracies for {\it NuSTAR}.

The best-fit centroids of the two Arp 299 sources are consistent with the positions of nuclei A and B, with
best-fit Gaussian $\sigma$ values were $\sim 4-5\arcsec$.  The southern serendipitous source extent was $\sim 2\arcsec$.  For all sources, any uncertainty in the PSF model (minimally due to unmodeled mirror - detector motion) is contributing to the fitted $\sigma$ values, and since the southern source is an isolated point source (i.e., the source is point-like in the \chandra image in Figure 3), this error is likely on the order of 2\arcsec.  For the Arp 299 nuclear regions, multiple sources are evident in the hard band \chandra image must be contributing (since they overlap in energy) to the two \nustar ``sources'' (i.e., modeled with just two Gaussians since the \chandra sources are effectively unresolved to {\it NuSTAR}).  

We  similarly fit the 10--20 keV image (using the same PSF models since the PSF is largely independent of energy) with the same three-Gaussian model.  Since the eastern source is not required in this case, we fixed the positions of the sources to the 6--10 keV fit values and the $\sigma$ of the eastern source to 2\arcsec\ to be effectively unresolved (as stated above, the amount of blurring necessary for the PSF as indicated by the unresolved southern source).  We then increased the normalization of the Gaussian representing the eastern X-ray emission of the galaxy until $C$ increased by 4.6.  This resulted in a 90\% confidence upper-limit of $\sim 60$ counts, or $\sim 4\%$ of the total 10--20 keV flux from Arp 299, that could be due to Arp 299-A and X-ray binaries.  As a consistency check we computed the number of counts expected from the \nustar image based on spectral modeling of extranuclear flux discussed below (the cut-off power-law model) in the 6--10 keV and 10--20 keV bands. This resulted in a factor of $\sim 2$ more counts than derived from the spatial analysis.  We take this to represent the systematic error in the analysis, such as errors in the background estimation and modeling all extranuclear flux as a single source.\footnote{Future work will use a more detailed modeling of the extended \nustar flux using the \chandra data as priors.}
Therefore we conservatively estimate the upper-limit of 10--20 keV flux extranuclear to Arp 299-B to be 10\%. 

\begin{table*}
\center
\caption{Spatial Fit Results for Arp 299 \nustar Images}\label{spattab}
\begin{tabular}{llcccc}
\hline
\hline
Source & Parameter & 6--10 keV Fit & Error & 10--20 keV Fit & Error  \\
\hline
West & R.A. & 11:28:30.45 & $-1.7$\arcsec, $+$1.8\arcsec & 11:28:30.45 & fixed \\
& Dec. & 58:33 39.9 & $-1.9$\arcsec, $+$1.6\arcsec & 58:33 39.9 & fixed \\ 
& $\sigma$ (\arcsec) & 4.1 & 3.9-4.3 & 4.5 &  4.4-4.6\\
& Counts & 938 & & 1615\\ 
East & R.A. & 11:28:33.10 & $-4.5$\arcsec, $+$5.1\arcsec & 11:28:33.10 & fixed \\
& Dec. &  58:33:46.8 & $-4.8$\arcsec, $+$4.5\arcsec & 58:33:46.8  & fixed \\
& $\sigma$ (\arcsec) & 4.0 & 3.3-4.5 & 2.0 & fixed\\ 
& Counts & 230  & & $<$60  \\
South & R.A. &11:28:29.22 & $-3.5$\arcsec, $+$3.4\arcsec & 11:28:28.97 & \\
& Dec. &58:31:58.6 & $-2.8$\arcsec,$+$3.5\arcsec & 58:31:59.8 & \\
& $\sigma$ (\arcsec) & 1.5 & 1.3-1.7 & 1.5& 1.4-1.7\\
& Counts & 199  & & 174\\\
Bgd. Model Counts & & 714 &  & 875\\
\hline
\end{tabular}
\tablecomments{These fits include the background model image generated by the code described in Wik et al. (2014).}
\end{table*}

\subsection{Spectral Results}
\subsubsection{The \nustar Spectra}
\begin{figure}
\epsfig{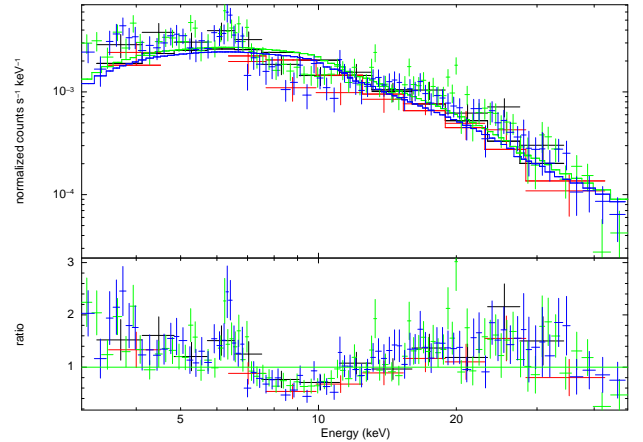}
\caption{\label{plfit} Simple (absorbed) power-law fit to the \nustar Arp 299 spectra.  The data is shown in the top panel with the best-fit model and the data/model ratio is shown in the bottom panel.  In both panels the points correspond to Observation 1 FPMA (black), FPMB (red); Observation 2 FPMA (green), FPMB (blue).}
\end{figure}

Figure \ref{plfit} shows the results of a single absorbed power-law model fit to the \nustar spectra (both detectors and both observations fit simultaneously), extracted from the 1\arcmin\ radius region shown in Figure \ref{chnuimgs}.  No spectral variability is detected and the normalizations of the two observations agree within 6\%. The fit is consistent with no absorption (note that \nustar is not sensitive to column densities less than $\sim 10^{21}\ \rm cm^{-2}$) and the best fit photon index $\Gamma$ was 0.8, albeit with $\chi^2$/dof=504/183, showing the fit is poor.
There is clear curvature detected in the spectrum, most notably due to Fe-K emission at 6.4--6.7 keV and a ``hump'' peaking around 20--30 keV. Both features are typical of highly obscured AGN \citep[e.g.,][Balokovic et al. 2014 submitted]{2013Natur.494..449R}, and require a more complex spectral model.  Note that the energy resolution of the \nustar detectors is $\sim 400$ eV FWHM \citep[independent of energy below 50 keV; ][]{Harrison:2013iq} and is therefore unable to cleanly resolve neutral (6.4 keV) and He-like (6.7 keV) Fe-K emission.

\subsubsection{The \chandra Spectrum}
\begin{figure}
\epsfig{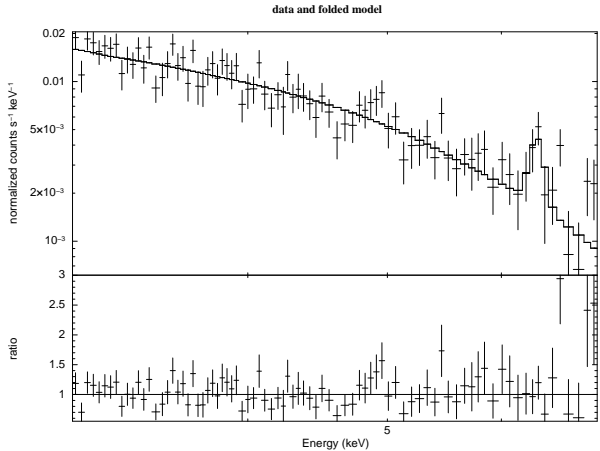}
\caption{\label{chonly} Power-law plus Gaussian fit to the 3--8 keV \chandra spectrum from a 1\arcmin\ region corresponding to the \nustar Arp 299 spectral source extraction region.}
\end{figure}

We fit the $3-8$ keV \chandra spectrum extracted from the same 1\arcmin\ region with a simple power-law.  In this case we included a Gaussian model to fit the Fe-K emission previously seen in Arp 299 by prior investigations.  Since more detailed analysis of the \chandra data will be presented in future work, our aim here is only to assess the extent to which a more complex model than a single power law is needed to fit the hard ($E>3$ keV) \ch continuum.  The power-law model provides an acceptable fit (Figure \ref{chonly}; $\chi^2$/dof = 80.6/77), showing that no thermal or other components contribute significantly in the 3--8 keV band.  The 2--10 keV flux from the fit is $1.1 \times 10^{-12} \ \rm\ ergs\ cm^{-2}\ s^{-1}$.  The power-law slope is 
$\Gamma = 1.9\ (1.6-2.2$; 90\% confidence)  and the rest-frame Fe-K line energy is 6.4 (6.3-6.6) keV, consistent with neutral Fe-K but with an ionized line likely contributing.

\subsubsection{Joint \chandra - \nustar Fits}
Given the complexity evident in the \nustar spectrum, we proceed with multi-component fits.  
The models we attempted were 1) power-law plus an absorbed cut-off power-law (hereafter the ``simple'' partial-covering model)  
and 2) a ``decoupled'' MYTorus model \citep{Murphy:2009p5869}, where the line-of-sight $N_{\rm H,Z}$ (MYTorusZ with inclination angle fixed at $90^{\circ}$) fit independently of the ``average'' $N_{\rm H,S}$ of material scattering photons into the line of sight (MYTorusS and MYTorusL with inclination angle fixed at $0^{\circ}$;  In our fits $N_{\rm H,Z}$ and $N_{\rm H,S}$ are consistent within the errors).  MyTorus is a model that self-consistency computes the transmitted, absorbed and reflected components of a power-law source surrounded by a toroidal absorber.  The tbabs absorption model in XSPEC was used for cold neutral absorption components in the partial covering model and a global absorption (i.e., due to absorption towards the Arp 299 system and absorption due to extranuclear gas) in the MYTorus model fit (while again the AGN absorption is handled self-consistently in the MYTorus model).

\begin{figure}
\epsfig{file=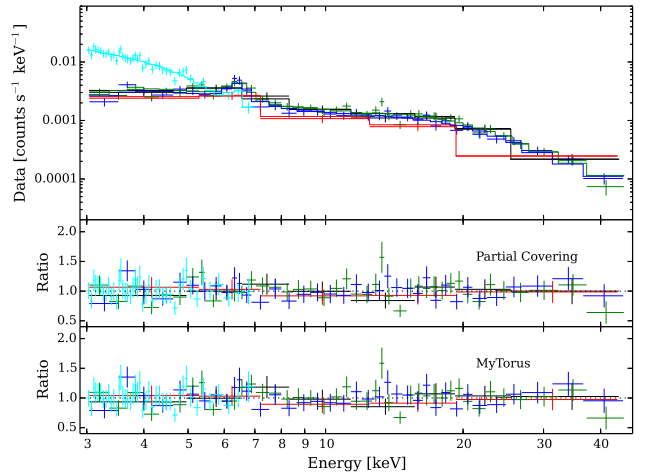,width=1.1\linewidth}
\caption{\label{specfits} Simultaneous \nustar + \chandra fits to Arp 299.  The top panel shows the data with the MYTorus model, and the bottom panels show the data/model ratio for the partial covering and MYTorus models as marked.}
\end{figure}

The simple partial-covering model does not include Compton scattering and therefore is not physical for systems like this with high column densities.  It also makes no assumptions about the geometry of the source but we included it to ease comparison to prior results in the literature. We allowed the power-law slopes in the partial covering model to vary independently since below 10~keV we know that both the nuclei and X-ray binaries are contributing, while above 10 keV the spectrum must be dominated by Arp 299-B alone.    
The MYTorus fit also includes an additional unabsorbed power-law to represent any AGN flux scattered around the ``torus'' and hence not accounted for by MYTorus.  The normalization of this scattered component relative to the intrinsic power-law was constrained by the fit to be extremely low, $< 10^{-3}$. 
The model parameters are given in Table \ref{specfitstab} and the best-fit spectra are plotted in Figure \ref{specfits}. 

Since the spatial analysis has shown that the 10--40 keV X-ray flux is concentrated on Arp 299-B, we associate the highly obscured AGN in these spectral fits with Arp 299-B, as first suspected from the \bepposax detection and neutral Fe-K localized to Arp 299-B.  In order to more precisely account for any flux external to Arp 299-B we fit the \chandra spectrum of the 1\arcmin\ region {\it excluding} the Arp 299-B nuclear source (radius of 2.5\arcsec) with a simple cut-off power-law model. This \chandra spectrum then represents the flux of X-ray binaries and Arp 299-A.  We fixed the cut-off energy at 7 keV to avoid over-predicting the count rate of the emission other than the Arp 299-B AGN above 10 keV.  The resulting fit parameters were $\Gamma$ = 1.6 (1.2-2.0) and a 2--10 keV flux of $8.1 \times 10^{-13} \rm \ ergs\ cm^{-2} \ s^{-1}$. This cut-off power-law model was added to the MYTorus fit since the MYTorus model pertains exclusively to AGN emission.  We set $\Gamma$ and the power-law normalization to be constrained to the range from the \chandra fit (the ``A-XRB'' component in Table \ref{specfitstab}). 

\begin{figure}
\epsfig{file=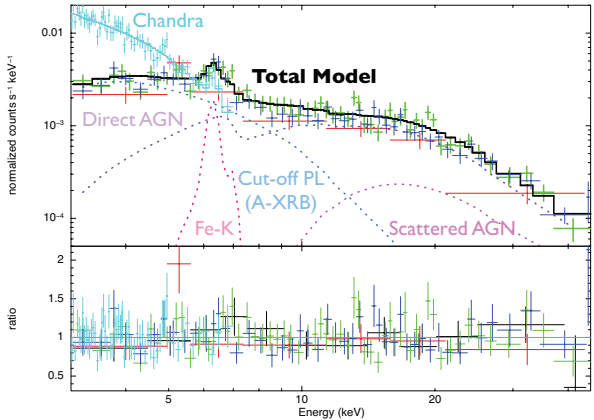,angle=90,width=1.0\linewidth}
\caption{\label{mytorusfit} Simultaneous \nustar + \chandra fits to Arp 299 with the MYTorus model, including a cut-off power-law to model flux due to sources other than Arp 299-B (e.g., Arp 299-A and X-ray binaries, noted as A-XRB), dominating above 10 keV.   The ``Direct AGN'' component is the line-of-sight AGN power-law component modified by the MYTorus $N_{\rm H,Z}$ model and the ``Scattered AGN" component is the $N_{\rm H,S}$ MYTorus model (reflection from the opposite side of the toroidal material often modeled using Compton reflection from a disk). The unabsorbed power-law constrained by the fit to have a low normalization is not shown.}
\end{figure}

Figure \ref{mytorusfit} shows the MYTorus fit with individual model components highlighted.  From this figure it is evident that the extranuclear cut-off power-law dominates below 10 keV and fits the \nustar and \chandra\ spectra well, and therefore the constraint on scattered, unabsorbed flux ($f$ in Table \ref{specfitstab} ) is tight.   Note that the MyTorus model assumes a (fixed) 60 degree opening angle for the torus, while for a LIRG such as Arp 299 the obscuration may be due in part to material associated with the starburst outside of the nuclear region.  In that case, a somewhat larger covering fraction may be appropriate. However, as can be seen in Figure \ref{mytorusfit}, above $\sim 6$ keV the directly absorbed emission of the AGN dominates the spectrum and the contribution of reflected or scattered emission is small.  Assuming a larger covering fraction therefore would not have a large impact.

As stated above, past observations have shown that both ionized and neutral Fe-K lines are present in Arp 299 \citep{2004ApJ...600..634B}.
 Not surprisingly, a single Gaussian included in the partial covering model resulted in an energy intermediate between neutral (6.4 keV) and He-like (6.7 keV) Fe-K.  
The MYTorus model 
inherently includes neutral Fe-K emission, and 
including an additional narrow ($\sigma$ fixed at 0.01 keV) Gaussian line results in $E = 6.64 (6.60-6.70)$ and a normalization of $2.4^{+1.7}_{-1.2} \times 10^{-6}$ photons s$^{-1}$ cm$^{-2}$, consistent with the ionized Fe-K flux reported in \citet{2004ApJ...600..634B} of 3.0 $\times 10^{-6}$ photons s$^{-1}$ cm$^{-2}$.  

\begin{table*}
\footnotesize
\center
\caption{\label{specfitstab}Chandra + NuSTAR Spectral Fits to Arp 299}
\begin{tabular}{cc}
\hline
Parameter & Value \\
\hline
\multicolumn{2}{c}{\bf{Partial Covering with Cut-off Power-law}}\\
$N_{\rm H} (10^{22} \rm cm^{-2})$ & 1.1 (0.6-1.6) \\
$\Gamma_1$ & 2.10 (1.86-2.12) \\
E (keV) & 6.49 (6.40-6.59) \\ % rest-frame
$\sigma$ (keV) & 0.22 (0.12-0.35)\\
Norm. (photons cm$^{-2}$ s$^{-1}$) & $7.3 (5.6-9.8) \times  10^{-6}$\\
EW (keV) & 0.64 (0.49-0.86) \\
$N_{\rm H,2} (10^{24} \rm cm^{-2})$ & 2.2 (0.7-3.4) \\
$\Gamma_2$ & 0.9 (-1.5 -- 2.0) \\
$E_{\rm cut}$ (keV) & 39 ($>$ 10) \\
$C$/dof & 843/861 \\
\multicolumn{2}{c}{\bf{Decoupled MyTorus plus Scattering and X-ray
  Binaries}}\\
$N_{\rm H} (10^{22} \rm cm^{-2})$ & 0.55 (0.04-1.07) \\
$\Gamma$ & 1.95 (1.54-1.96) \\
$N_{\rm H,Z} (10^{22} \rm cm^{-2})$ & 350 ($>$330) \\
$N_{\rm H,S} (10^{22} \rm cm^{-2})$\tablenotemark{a} & 410 (330-500) \\
$A_S$\tablenotemark{b} & 0.39 0.34-0.46) \\
$A_L$\tablenotemark{b} & 0.05 (0.02-1.2) \\
A-XRB $\Gamma$\tablenotemark{c} & 1.80 (1.77-1.83) \\
A-XRB $E_{\rm cutoff}$ (keV) & 7.0 (fixed) \\
 A-XRB Norm (photons cm$^{-2}$ s$^{-1}$ at 1 keV) & $5.4 (5.1-5.6) \times 10^{-4}$\\
$f_{\rm scatt}$\tablenotemark{d} & $0 (<1.1 \times 10^{-3})$ \\
C/dof & 849/861 \\
\hline
\end{tabular}
\tablenotetext{1}{$N_{\rm H,S}$ is the column density for both the MyTorus reflected/scattered component  (MYTorusS) and line component (MYTorusL).}
\tablenotetext{2}{$A_S$ and $A_L$ are normalization factors applied to
  the MYTorus reflected and line components, respectively.}
\tablenotetext{3}{A-XRB $\Gamma$, Norm are the power-law slope and normalization of a
cut-off power-law included to account for the unresolved X-ray nuclear
emission, i.e., due to the nucleus of Arp 299-A and X-ray binaries.  The bounds of these parameters were constrained to be
consistent with the {\it Chandra} 3-7 keV spectrum excluding Arp 299-B when
fit with the cut-off power-law model.}
\tablenotetext{4}{``Scattering'' factor applied to the intrinsic power-law
  (i.e., photon index tied to the MYTorus power-law) to represent flux
  scattered {\it around} the toroidal obscuration from
  highly ionized material.}
\end{table*}
%\end{singlespace}

\section{Discussion}
\subsection{The AGN Arp 299-B}
These observations represent the first detection of Arp 299 out to $\sim45$ keV, and the first unambiguous determination that the $E>10$ keV flux is predominantly due to Arp 299-B alone.  The best-fit $N_{\rm H}=1-3 \times 10^{24}\ \rm cm^{-2}$  from the simple partial-covering model, and $N_{\rm H} \sim 4 \times 10^{24} \rm \ cm^{-2}$ from the decoupled MYTorus model, which is more physical since this model is self-consistent in its treatment of absorption and scattering. The observed 2--10 keV and 10--30 keV fluxes for Arp 299 from the MYTorus model are $1.1 \times 10^{-12} \rm \ ergs \ cm^{-2} \ s^{-1}$ and $3.5\times 10^{-12} \rm \ ergs \ cm^{-2} \ s^{-1}$, respectively.  \bepposax detected Arp 299 with similar spectral parameters, namely $\Gamma = 1.8$ and $N_{\rm H} = 2.5 \times 10^{24} \rm \ cm^{-2}$ 
\citep{DellaCeca:2002hc}.
However the 2--10 keV and 10--100 keV fluxes detected by \bepposax were $1.1 \times 10^{-12}$ and $3.2 \times 10^{-11} \rm \ ergs \ cm^{-2} \ s^{-1}$.  The 10--30 keV flux from the \bepposax fit is $7.1 \times 10^{-12} \rm \ ergs \ cm^{-2} \ s^{-1}$.
Therefore the observed 10--30 keV flux detected by \nustar is a factor of $\sim 2$ lower than the \bepposax detection. Given the consistency in the measured column densities between the \nustar and \bepposax fits, any variability is likely to be mostly due to changes in intrinsic flux and not absorption.  Note that the serendipitous source to the south of Arp 299 is only contributing $\sim 10\%$ of the counts in the 10--20 keV band at the current time but it may be variable.

\citet{Koss:2013uj}
reported a $\sim 3\sigma$ detection of Arp 299 by the {\it Swift} BAT with a 14-195 keV luminosity of 1.1 (0.7-2.0) $\times 10^{42}$ ergs s$^{-1}$.  
\citet{Koss:2013uj}
 assumed the Crab spectrum (generally used in the calibration of BAT fluxes), namely a power-law spectrum with $\Gamma=2.0$, which gives a 10-30~keV flux of 1.9 (1.1-3.4) $\times 10^{-12}$ ergs cm$^{-2}$  s$^{-1}$, consistent with the \nustar 10--30 keV flux within the errors.   For completeness we derived the BAT count rate from the reported luminosity in Koss et al. and used the BAT spectral response with the best-fitting MyTorus model to determine the BAT flux.  This resulted in a 10--30 keV flux of $1.4 \times 10^{-12}$ ergs cm$^{-2}$  s$^{-1}$.  Therefore, the average 10-30 keV flux of Arp 299 from 2004-2010 was consistent with the \nustar flux given the error on the BAT flux but was possibly a factor of $\sim 2$ lower. 
The neutral Fe-K flux implied by the MYTorus fit is $\sim 3.5 \times 10^{-6}$ photons cm$^{-2}$ s$^{-1}$.  The neutral Fe-K flux reported in \citet{2004ApJ...600..634B} was $1.9 (0.6-3.1) \times 10^{-6}$ photons cm$^{-2}$ s$^{-1}$ and is therefore consistent.  
However, if the neutral Fe-K line in Arp 299 was due to material more than $\sim 12$ light years from the nucleus then the line would not be expected to respond to changes in the continuum flux on the $\sim 12$ year time scale between the \bepposax and \xmmn observations, both in 2001, and the \nustar observation in 2013.  

The intrinsic (absorption corrected) X-ray luminosity ($L_X$) in the 2--10 and 10--30 keV bands bands are $1.5 \times 10^{43}$ and $1.2\times 10^{43}$ ergs s$^{-1}$, respectively.  From an analysis of the mid-IR spectra, 
\citet{AlonsoHerrero:2013ct} estimate the bolometric luminosity of Arp 299-B to be $3 \times 10^{44}$  ergs s$^{-1}$.  Therefore our results are consistent with $\sim 10\%$ of the bolometric flux being emitted in X-rays, which is consistent with typical AGN\citep{Vasudevan:2007p4739}.   Arp 299-B has a velocity dispersion of 144 km~s$^{-1}$
\citep{Ho:2009de}, giving a central black hole mass of $\sim 2 \times 10^{7}\rm \ M_{\odot}$ based on the bulge velocity dispersion-BH mass relation (note that since Arp 299 is a merging system the observed velocity dispersion may not follow this correlation) in \citet{Graham:2011hc}. 
The implied Eddington luminosity is then $\sim 3\times 10^{45}$ ergs s$^{-1}$.  Therefore $L_{\rm X}/L_{\rm Edd} \sim 5-10 \%$ which is also consistent with the $L_{\rm X}/L_{\rm Edd}$ distribution observed in quasars 
\citep{Kelly:2010fk}.
If the \bepposax hard X-ray flux was due to variability of Arp 299-B, then at that time the AGN was accreting at $\sim 10-20\%\ L_{\rm Edd}$.  

\subsection{Arp 299-A and X-ray Binaries}
Since both galaxies in a merger likely have nuclear supermassive black holes, it would be expected that the merging process may fuel both black holes if the merging galaxies are gas-rich 
\citep{Yu:2011gz}.
 However, this is expected to result in detectable binary AGN only rarely since this fueling occurs during the latest stages of the merger and is episodic
\citep{VanWassenhove:2012ci}.
\citet{AlonsoHerrero:2013ct}, \citet{2004ApJ...600..634B} and \citet{2010A&A...519L...5P} argue that Arp 299-A also harbors an AGN on the basis of its mid-IR spectrum, the detection of ionized Fe-K emission and a flat-spectrum radio source, respectively.  \citet{AlonsoHerrero:2013ct} 
 argue that Arp 299-A is heavily obscured based on SED fitting, which also implies, based on a poor fit to the clumpy torus model, that the obscuration is likely to be spherical.  Here we constrain the contribution of Arp 299-A above 10 keV to be less than $\sim 10\%$ of the total emission, implying 
$L_{10-30\rm \ keV} < 1.2 \times 10^{42}$ ergs s$^{-1}$.  
This would imply a low-luminosity AGN and/or high obscuration.

As noted in \citet{2004ApJ...600..634B}, the lack of a strong neutral line from Arp 299-A implies that the nucleus is not heavily obscured.   \citet{2004ApJ...600..634B} argue that $\sim 20\%$ of the ionized Fe-K in Arp 299-A can be associated with star formation, i.e., X-ray binaries and hot ISM, based on scaling the NGC 253 line flux by the star-formation rate (SFR) of Arp 299-A.  We can similarly compare the ionized Fe-K in Arp 299-A to that in Arp 220.  The ionized Fe-K EW reported for Arp 299-A in \citet{2004ApJ...600..634B} is $\sim 0.8$ keV, while the Fe-K EW for Arp 220 reported in \citet{Teng:2009p8908} is $0.42^{+0.54}_{-0.32}$ keV.  Following \citet{2004ApJ...600..634B} and taking the 1.4 Ghz luminosity as a proxy for the FIR luminosity (and hence a star formation rate indicator),
\citet{AlonsoHerrero:2000fr} argues that $\sim 50\%$ of the FIR emission from Arp 299 is associated with Arp 299-A, or $L_{IR} = 9.6 \times 10^{44}$ ergs s$^{-1}$.  This is a factor of $\sim 6$ lower than the IR luminosity of Arp 220, implying that $\sim 10\%$ of the ionized Fe-K is due to star formation.
However clearly a larger sample of starburst galaxies with detected ionized Fe-K is needed for a meaningful comparison.   Therefore the current evidence (including the lack of strong X-ray emission above 10 keV)  implies that Arp 299-A harbors a low-luminosity AGN and a starburst, both of which are contributing to the Fe-K emission and hence the energy budget, although the starburst is dominating both the total SED \citep{AlonsoHerrero:2013ct} and the 3-10 keV X-ray emission from Arp 299-A.

As discussed above, the mean spectrum of the extranuclear sources in Arp 299 must be cut-off above 10 keV to avoid over-predicting the observed $E>10$ keV \nustar flux.  The X-ray binaries in NGC 253 contributing to the observed \nustar flux are also soft, $\Gamma \sim 3$ at $E> 10$ keV.  To date observations of ULXs by \nustar 
\citep{Lehmer:2013bs,2013ApJ...779..148W,Bachetti:2013fd,Rana:2014tr,Walton:2014ta}
are also showing these sources to be soft or cut-off at $E>10$ keV.   Therefore a consistent picture is that all high luminosity XRBs (i.e., $L_X > 10^{38}$ ergs s$^{-1}$) in starburst galaxies have on average soft spectra above 10 keV, similar to ULXs.  This will be further investigated when the full starburst galaxy sample is observed with {\it NuSTAR}.

\section{Summary}
We have presented the first results of observing the Arp 299 merger system with {\it NuSTAR}.  Our main conclusion is that the hard X-ray emission ($E>10$ keV) is dominated by only one of the two nuclei, Arp 299-B.  The \nustar and {\it Swift} BAT $10-30$ keV fluxes for Arp 299 are a factor of $\sim 2$ lower than the \bepposax flux although both the \nustar and \bepposax spectra implied similar column densities, $N_{\rm H} \sim 2-5 \times 10^{24} \rm \ cm^{-2}$.  Therefore, if Arp 299-B varied since the \bepposax observation the variability was likely to be due to a change in intrinsic luminosity rather than a change in absorption.  

Arp 299-A and all other sources of hard X-ray emission, namely any AGN in Arp 299-A and the X-ray binary population of Arp 299, are contributing at most 10\% of the $E>10$ keV flux from Arp 299.  This implies that the Arp 299-A nucleus and the X-ray binaries must be soft or cut-off above $\sim 10$ keV.  X-ray binaries in the starburst galaxies NGC 253 are similarly soft above 10 keV \citep{Lehmer:2013bs,Wik:2014wt} as are ULXs observed to date by \nustar \citep{2013ApJ...779..148W,Bachetti:2013fd,Rana:2014tr,Walton:2014ta}.  The planned observations of additional starburst galaxies (M83, NGC 3310 and NGC 3256) with \nustar should show if this behavior is generally true for luminous extragalactic X-ray binaries.  Future observations of Arp 299 with imaging hard X-ray telescopes would help to determine if the nuclei tend are variable.

\acknowledgements
This research has made use of data obtained with the
\nustar mission, a project led by the California Institute of
Technology (Caltech), managed by the Jet Propulsion Laboratory
(JPL) and funded by NASA.  The scientific results reported in
this article are based in part on observations made by
the {\it Chandra X-ray Observatory}. We thank the \nustar Operations,
Software and Calibration teams for support with the
execution and analysis of these observations. This research
has made use of the \nustar Data Analysis Software (NUSTARDAS),
jointly developed by the ASI Science Data Center
(ASDC, Italy) and Caltech (USA). We also made use of the
NASA/IPAC Extragalactic Database (NED) and NASA's Astrophysics Data System. LB receive financial supports  from the European Commission Seventh Framework Programme (FP7/2007-2013) under grant agreement no 267251â Astronomy Fellowships in Italyâ (AstroFIt). A. Zezas acknowledges partial support by NASA grant NNX12AN05G, and Chandra grant GO3-14124X.  We thank the anonymous referee for suggestions that improved this paper.

\bibliography{ms.bbl}

\begin{thebibliography}{}
\expandafter\ifx\csname natexlab\endcsname\relax\def\natexlab#1{#1}\fi

\bibitem[{Alonso-Herrero {et~al.}(2000)Alonso-Herrero, Rieke, Rieke, \&
  Scoville}]{AlonsoHerrero:2000fr}
Alonso-Herrero, A., Rieke, G.~H., Rieke, M.~J., \& Scoville, N.~Z. 2000, ApJ,
  532, 845

\bibitem[{Alonso-Herrero {et~al.}(2013)Alonso-Herrero, Roche, Esquej,
  Gonzalez-Mart{\'\i}n, Pereira-Santaella, Ramos~Almeida, Levenson, Packham,
  Asensio~Ramos, Mason, Rodr{\'\i}guez~Espinosa, Alvarez, Colina, Aretxaga,
  Diaz-Santos, Perlman, \& Telesco}]{AlonsoHerrero:2013ct}
Alonso-Herrero, A., Roche, P.~F., Esquej, P., {et~al.} 2013, ApJ, 779, L14

\bibitem[{Bachetti {et~al.}(2013)Bachetti, Rana, Walton, Barret, Harrison,
  Boggs, Christensen, Craig, Fabian, F{\"u}rst, Grefenstette, Hailey,
  Hornschemeier, Madsen, Miller, Ptak, Stern, Webb, \& Zhang}]{Bachetti:2013fd}
Bachetti, M., Rana, V., Walton, D.~J., {et~al.} 2013, ApJ, 778, 163

\bibitem[{Ballo {et~al.}(2004)Ballo, Braito, Della~Ceca, Maraschi, Tavecchio,
  \& Dadina}]{2004ApJ...600..634B}
Ballo, L., Braito, V., Della~Ceca, R., {et~al.} 2004, ApJ, 600, 634

\bibitem[{Comerford {et~al.}(2009)Comerford, Gerke, Newman, Davis, Yan, Cooper,
  Faber, Koo, Coil, Rosario, \& Dutton}]{Comerford:2009fo}
Comerford, J.~M., Gerke, B.~F., Newman, J.~A., {et~al.} 2009, ApJ, 698, 956

\bibitem[{Cotini {et~al.}(2013)Cotini, Ripamonti, Caccianiga, Colpi,
  Della~Ceca, Mapelli, Severgnini, \& Segreto}]{Cotini:2013fw}
Cotini, S., Ripamonti, E., Caccianiga, A., {et~al.} 2013, Monthly Notices RAS,
  431, 2661

\bibitem[{Della~Ceca {et~al.}(2002)Della~Ceca, Ballo, Tavecchio, Maraschi,
  Petrucci, Bassani, Cappi, Dadina, Franceschini, Malaguti, Palumbo, \&
  Persic}]{DellaCeca:2002hc}
Della~Ceca, R., Ballo, L., Tavecchio, F., {et~al.} 2002, ApJ, 581, L9

\bibitem[{Fabbiano {et~al.}(2011)Fabbiano, Wang, Elvis, \&
  Risaliti}]{Fabbiano:2011fo}
Fabbiano, G., Wang, J., Elvis, M., \& Risaliti, G. 2011, Nature, 477, 431

\bibitem[{Fabian {et~al.}(1998)Fabian, Barcons, Almaini, \&
  Iwasawa}]{Fabian:1998es}
Fabian, A.~C., Barcons, X., Almaini, O., \& Iwasawa, K. 1998, Monthly Notices
  RAS, 297, L11

\bibitem[{Graham {et~al.}(2011)Graham, Onken, Athanassoula, \&
  Combes}]{Graham:2011hc}
Graham, A.~W., Onken, C.~A., Athanassoula, E., \& Combes, F. 2011, Monthly
  Notices RAS, 412, 2211

\bibitem[{Harrison {et~al.}(2013)Harrison, Craig, Christensen, Hailey, Zhang,
  Boggs, Stern, Cook, Forster, Giommi, Grefenstette, Kim, Kitaguchi, Koglin,
  Madsen, Mao, Miyasaka, Mori, Perri, Pivovaroff, Puccetti, Rana, Westergaard,
  Willis, Zoglauer, An, Bachetti, Barriere, Bellm, Bhalerao, Brejnholt, Fuerst,
  Liebe, Markwardt, Nynka, Vogel, Walton, Wik, Alexander, Cominsky,
  Hornschemeier, Hornstrup, Kaspi, Madejski, Matt, Molendi, Smith, Tomsick,
  Ajello, Ballantyne, Balokovi{\'c}, Barret, Bauer, Blandford, Brandt,
  Brenneman, Chiang, Chakrabarty, Chenevez, Comastri, Dufour, Elvis, Fabian,
  Farrah, Fryer, Gotthelf, Grindlay, Helfand, Krivonos, Meier, Miller,
  Natalucci, Ogle, Ofek, Ptak, Reynolds, Rigby, Tagliaferri, Thorsett,
  Treister, \& Urry}]{Harrison:2013iq}
Harrison, F.~A., Craig, W.~W., Christensen, F.~E., {et~al.} 2013, ApJ, 770, 103

\bibitem[{Ho {et~al.}(2009)Ho, Greene, Filippenko, \& Sargent}]{Ho:2009de}
Ho, L.~C., Greene, J.~E., Filippenko, A.~V., \& Sargent, W. L.~W. 2009, The
  Astrophysical Journal Supplement Series, 183, 1

\bibitem[{Hopkins {et~al.}(2006)Hopkins, Hernquist, Cox, Di~Matteo, Robertson,
  \& Springel}]{Hopkins:2006p2661}
Hopkins, P.~F., Hernquist, L., Cox, T.~J., {et~al.} 2006, The Astrophysical
  Journal Supplement Series, 163, 1

\bibitem[{Iwasawa {et~al.}(2009)Iwasawa, Sanders, Evans, Mazzarella, Armus, \&
  Surace}]{Iwasawa:2009p4997}
Iwasawa, K., Sanders, D.~B., Evans, A.~S., {et~al.} 2009, The Astrophysical
  Journal Letters, 695, L103

\bibitem[{Kelly {et~al.}(2010)Kelly, Vestergaard, Fan, Hopkins, Hernquist, \&
  Siemiginowska}]{Kelly:2010fk}
Kelly, B.~C., Vestergaard, M., Fan, X., {et~al.} 2010, ApJ, 719, 1315

\bibitem[{Komossa {et~al.}(2003)Komossa, Burwitz, Hasinger, Predehl, Kaastra,
  \& Ikebe}]{Komossa:2002js}
Komossa, S., Burwitz, V., Hasinger, G., {et~al.} 2003, ApJ, 582, L15

\bibitem[{Koss {et~al.}(2013)Koss, Mushotzky, Baumgartner, Veilleux, Tueller,
  Markwardt, \& Casey}]{Koss:2013uj}
Koss, M., Mushotzky, R., Baumgartner, W., {et~al.} 2013, ApJ, 765, L26

\bibitem[{Lacki \& Thompson(2013)}]{Lacki:2013hz}
Lacki, B.~C., \& Thompson, T.~A. 2013, ApJ, 762, 29

\bibitem[{Lehmer {et~al.}(2013{\natexlab{a}})Lehmer, Lucy, Alexander, Best,
  Geach, Harrison, Hornschemeier, Matsuda, Mullaney, Smail, Sobral, \&
  Swinbank}]{Lehmer:2013bs}
Lehmer, B.~D., Lucy, A.~B., Alexander, D.~M., {et~al.} 2013{\natexlab{a}}, ApJ,
  765, 87

\bibitem[{Lehmer {et~al.}(2013{\natexlab{b}})Lehmer, Wik, Hornschemeier, Ptak,
  Antoniou, Argo, Bechtol, Boggs, Christensen, Craig, Hailey, Harrison,
  Krivonos, Leyder, Maccarone, Stern, Venters, Zezas, \& Zhang}]{Lehmer:2013bk}
Lehmer, B.~D., Wik, D.~R., Hornschemeier, A.~E., {et~al.} 2013{\natexlab{b}},
  ApJ, 771, 134

\bibitem[{Liu {et~al.}(2013)Liu, Civano, Shen, Green, Greene, \&
  Strauss}]{Liu:2012jw}
Liu, X., Civano, F., Shen, Y., {et~al.} 2013, ApJ, 762, 110

\bibitem[{Mitsuishi {et~al.}(2011)Mitsuishi, Yamasaki, \&
  Takei}]{Mitsuishi:2011cp}
Mitsuishi, I., Yamasaki, N.~Y., \& Takei, Y. 2011, ApJ, 742, L31

\bibitem[{Murphy \& Yaqoob(2009)}]{Murphy:2009p5869}
Murphy, K.~D., \& Yaqoob, T. 2009, Monthly Notices RAS, 889

\bibitem[{P{\'e}rez-Torres {et~al.}(2010)P{\'e}rez-Torres, Alberdi,
  Romero-Canizales, \& Bondi}]{2010A&A...519L...5P}
P{\'e}rez-Torres, M.~A., Alberdi, A., Romero-Canizales, C., \& Bondi, M. 2010,
  A{\&}A, 519, L5

\bibitem[{Rana {et~al.}(2014)Rana, Harrison, Bachetti, Walton, F{\"u}rst,
  Barret, Miller, Fabian, Boggs, Christensen, Craig, Grefenstette, Hailey,
  Madsen, Ptak, Stern, Webb, \& Zhang}]{Rana:2014tr}
Rana, V., Harrison, F.~A., Bachetti, M., {et~al.} 2014, eprint arXiv:1401.4637

\bibitem[{Risaliti {et~al.}(2013)Risaliti, Harrison, Madsen, Walton, Boggs,
  Christensen, Craig, Grefenstette, Hailey, Nardini, Stern, \&
  Zhang}]{2013Natur.494..449R}
Risaliti, G., Harrison, F.~A., Madsen, K.~K., {et~al.} 2013, Nature, 494, 449

\bibitem[{Satyapal {et~al.}(2014)Satyapal, Ellison, McAlpine, Hickox, Patton,
  \& Mendel}]{2014arXiv1403.7531S}
Satyapal, S., Ellison, S.~L., McAlpine, W., {et~al.} 2014, arXiv, 7531

\bibitem[{Teng {et~al.}(2009)Teng, Veilleux, Anabuki, Dermer, Gallo, Nakagawa,
  Reynolds, Sanders, Terashima, \& Wilson}]{Teng:2009p8908}
Teng, S.~H., Veilleux, S., Anabuki, N., {et~al.} 2009, ApJ, 691, 261

\bibitem[{Van~Wassenhove {et~al.}(2012)Van~Wassenhove, Volonteri, Mayer, Dotti,
  Bellovary, \& Callegari}]{VanWassenhove:2012ci}
Van~Wassenhove, S., Volonteri, M., Mayer, L., {et~al.} 2012, ApJ, 748, L7

\bibitem[{Vasudevan \& Fabian(2007)}]{Vasudevan:2007p4739}
Vasudevan, R.~V., \& Fabian, A.~C. 2007, Monthly Notices RAS, 381, 1235

\bibitem[{Walton {et~al.}(2013)Walton, Fuerst, Harrison, Stern, Bachetti,
  Barret, Bauer, Boggs, Christensen, Craig, Fabian, Grefenstette, Hailey,
  Madsen, Miller, Ptak, Rana, Webb, \& Zhang}]{2013ApJ...779..148W}
Walton, D.~J., Fuerst, F., Harrison, F., {et~al.} 2013, ApJ, 779, 148

\bibitem[{Walton {et~al.}(2014)Walton, Harrison, Grefenstette, Miller,
  Bachetti, Barret, Boggs, Christensen, Craig, Fabian, Fuerst, Hailey, Madsen,
  Parker, Ptak, Rana, Stern, Webb, \& Zhang}]{Walton:2014ta}
Walton, D.~J., Harrison, F.~A., Grefenstette, B.~W., {et~al.} 2014, eprint
  arXiv:1402.2992

\bibitem[{{Wik} {et~al.}(2014{\natexlab{a}}){Wik}, {Hornstrup}, {Molendi},
  {Madejski}, {Harrison}, {Zoglauer}, {Grefenstette}, {Gastaldello}, {Madsen},
  {Westergaard}, {Ferreira}, {Kitaguchi}, {Pedersen}, {Boggs}, {Christensen},
  {Craig}, {Hailey}, {Stern}, \& {Zhang}}]{Wik:2014vv}
{Wik}, D.~R., {Hornstrup}, A., {Molendi}, S., {et~al.} 2014{\natexlab{a}},
  \apj, 792, 48

\bibitem[{{Wik} {et~al.}(2014{\natexlab{b}}){Wik}, {Lehmer}, {Hornschemeier},
  {Yukita}, {Ptak}, {Zezas}, {Antoniou}, {Argo}, {Bechtol}, {Boggs},
  {Christensen}, {Craig}, {Hailey}, {Harrison}, {Krivonos}, {Maccarone},
  {Stern}, {Venters}, \& {Zhang}}]{Wik:2014wt}
{Wik}, D.~R., {Lehmer}, B.~D., {Hornschemeier}, A.~E., {et~al.}
  2014{\natexlab{b}}, \apj, 797, 79

\bibitem[{Yu {et~al.}(2011)Yu, Lu, Mohayaee, \& Colin}]{Yu:2011gz}
Yu, Q., Lu, Y., Mohayaee, R., \& Colin, J. 2011, ApJ, 738, 92

\bibitem[{Zezas {et~al.}(2003)Zezas, Ward, \& Murray}]{2003ApJ...594L..31Z}
Zezas, A., Ward, M.~J., \& Murray, S.~S. 2003, ApJ, 594, L31

\end{thebibliography}
\end{document}